\documentclass[twoside]{dis07}
\usepackage[latin1]{inputenc}
\usepackage[dvips]{graphicx,epsfig,color}
\usepackage{wrapfig,rotating}
\usepackage{amssymb,amsmath,array}

\newcommand{\abbrev}{\small}
\newcommand{\dred}{{\abbrev DRED}}
\newcommand{\dreg}{{\abbrev DREG}}
\newcommand{\mudec}{\mu_{\rm dec}}
\newcommand{\susy}{{\abbrev SUSY}}
\newcommand{\qcd}{{\abbrev QCD}}
\newcommand{\drbarmath}{\mbox{$\overline{\scriptstyle\mathrm{ DR}}$}}
\newcommand{\drbar}{\overline{{\mbox{\abbrev DR}}}}
\newcommand{\msbar}{\overline{{\mbox{\abbrev MS}}}}
\newcommand{\asMS}{\bar\alpha_s}
\newcommand{\asDR}{\alpha_s}
\newcommand{\aeDR}{\alpha_e}
\newcommand{\dd}{{\rm d}}
\newcommand{\ep}{\epsilon}
\newcommand{\epscalar}{$\epsilon$-scalar}
\newcommand{\eqn}[1]{Eq.\,(\ref{#1})}

\begin{document}
\title{Dimensional Reduction applied to QCD at higher orders}

\author{Robert Harlander$^1$, Philipp Kant$^2$, Luminita Mihaila$^2$,
  Matthias Steinhauser$^2$
%
\thanks{Work supported by the DFG through SFB/TR~9 and HA~2990/3-1.} 
\\[-10em]
  \hspace{16em}SFB/CPP-07-32, TTP/07-13, WUB/07-05 -- June 2007\\[2em]
    \vspace{5.5em}
%
\vspace{.3cm}\\
%
1- Fachbereich C - Bergische Universit\"at Wuppertal \\
42097 Wuppertal - Germany
%
\vspace{.1cm}\\
2- Theoretische Teilchenphysik - Universit\"at Karlsruhe \\
76128 Karlsruhe - Germany\\
}

\maketitle

\begin{abstract}
Recent developments in higher order calculations within the framework of
Dimensional Reduction, the preferred regularization scheme for
supersymmetric theories, are reported on. Special emphasis is put on the
treatment of evanescent couplings, the equivalence to Dimensional
Regularization, and the evaluation of $\alpha_s(M_{\rm GUT})$ from
$\alpha_s(M_{\rm Z})$. 
\end{abstract}



\section{Dimensional Reduction}


Dimensional Regularization (\dreg{})~\cite{'tHooft:fi} has proven
extremely successful for the evaluation of higher order corrections in
quantum field theory, mostly because it preserves gauge invariance and
thus does not interfere with the renormalizability of the Standard Model
or \qcd{}. Many techniques for evaluating Feynman diagrams have been
developed within the framework of \dreg{}, so perturbation theory
heavily relies upon the validity of this regularization method.

Applied to \susy{} theories, however, one faces the
problem of explicit \susy{} breaking by the need to assign different
numbers of degrees of freedom to spin-1 and spin-1/2 fields. A
manifestation of this \susy{} breaking is that \susy{} relations of
couplings no longer hold at higher orders. For example, while \susy{}
requires equality for the quark-quark-gluon and the squark-quark-gluino
couplings $g$ and $\hat g$ at all energy scales, one finds that their
renormalization constants differ. In fact, it is $Z_g = (1 +
\delta_{\hat g})\, Z_{\hat g}$, and thus
\begin{equation}
\begin{split}
\hat g = (1+\delta_{\hat g}) g\,,
\end{split}
\end{equation}
where $\delta_{\hat g} = \alpha_s/(3\pi)$~\cite{Martin:1993yx}. In
effect, the number of renormalization constants in \susy{} becomes
rather large when calculations are done in \dreg{}.

As a way out, it was suggested to use Dimensional Reduction (\dred{}) as
a regularization procedure for \susy{}
theories~\cite{Siegel:1979wq}. Formally, this means that space-time is
compactified to $D=4-2\ep$ dimensions ($\ep>0$), while the vector fields
are kept four-dimensional.  As an example, consider the electron-photon
vertex, which in \dred{} becomes
\begin{equation}
\begin{split}
\bar\psi \gamma_\mu\psi A^\mu = 
\bar\psi \gamma_\mu\psi \hat A^\mu +
\bar\psi \gamma_\mu\psi \tilde A^\mu
= \bar\psi \hat\gamma_\mu\psi \hat A^\mu +
\bar\psi \tilde\gamma_\mu\psi \tilde A^\mu\,,
\label{eq::psipsiA}
\end{split}
\end{equation}
where $\hat A^\mu$ and $\tilde A^\mu$ denote the $D$ and the
$2\ep$-dimensional component of the vector field $A^\mu$. $\tilde A^\mu$
is also called the \epscalar{}.  Traces over the $D$- and
$2\ep$-dimensional $\gamma$-matrices can be evaluated using
\begin{equation}
\begin{split}
\{\gamma^\mu,\gamma^\nu\} = 2g_{\mu\nu}\qquad\mbox{and}\qquad
\{\hat\gamma^\mu,\tilde\gamma^\nu\} = 0\,.
\end{split}
\end{equation}
Thus, perturbative calculations in \dred{} require to introduce
additional fields (\epscalar{}s) and an extra set of $\gamma$-matrices.
Once the algebraic part of the evaluation of a Feynman amplitude is
done, the tools developped for \dreg{} can be applied without further
modification.

\section{Evanescent couplings}

In a \susy{} theory, the relation $A_\mu=\hat A_\mu+\tilde A_\mu$ is
essential: it is $A_\mu$ that is part of a super-multiplet, while $\hat
A_\mu$ and $\tilde A_\mu$ are introduced for purely technical
reasons. Therefore, this relation must hold also at higher orders of
perturbation theory. 

This is not necessarily the case in a non-\susy{} theory. Since $\tilde
A_\mu$ transforms like a scalar under gauge transformations (thus the
name ``\epscalar''), there is no symmetry to ensure that the $\tilde
A_\mu$-couplings renormalize in the same way as the corresponding $\hat
A_\mu$-couplings. Thus, when applying \dred{} to \qcd{}, we have to
introduce two different couplings for the quark-gluon vertex, for
example:
\begin{equation}
\begin{split}
g_s A_\mu\bar\psi\gamma^\mu\psi\to
\hat g_s \hat A_\mu\bar\psi\hat \gamma^\mu\psi+
\tilde g_s \tilde A_\mu\bar\psi\tilde\gamma^\mu\psi\,.
\end{split}
\end{equation}
In order to be consistent with our journal
papers~\cite{Harlander:2006rj,Harlander:2006xq,Harlander:phen}\footnote{%
  There is a misprint in Eq.\,(18) of Ref.\,\cite{Harlander:2006xq}:
  the term $-25\,n_f^2/72$ should read $-25\zeta_3/72$.}
let us define
\begin{equation}
\begin{split}
\asDR{} &= \frac{\hat g_s^2}{4\pi}\,,\qquad
\alpha_e = \frac{\tilde g_s^2}{4\pi}\,,
\end{split}
\end{equation}
where $\alpha_e$ is called ``evanescent coupling''.  Only at tree-level
can one require that $\asDR{}=\alpha_e$.  Higher orders lead to an
energy dependece of the (minimally subtracted) couplings, governed by
the {\abbrev RGE}s\footnote{In fact, there are several evanescent
couplings in {\scriptsize QCD}; however, for the sake of the argument,
it is sufficient to consider only $\alpha_e$ here.}
\begin{equation}
\begin{split}
\mu^2\frac{\dd}{\dd\mu^2}\asDR{} =
\beta_s^{\drbarmath}(\asDR{},\alpha_e)\,,\qquad\quad
\mu^2\frac{\dd}{\dd\mu^2}\alpha_e = \beta_e(\asDR{},\alpha_e)\,.
\label{eq::rges}
\end{split}
\end{equation}
The $\beta$-functions have been evaluated in
Ref.\,\cite{Harlander:2006rj} through three loops, and
$\beta_s^{\drbarmath}$ is even known to four-loop
order~\cite{Harlander:2006xq}. Indeed it turns out that
$\beta_s^{\drbarmath}\neq \beta_e$ in standard \qcd{} already at
one-loop level. The condition $\asDR{}=\alpha_e$ can therefore be
implemented only at one particular value of $\mu^2$.

If $\drbar$ (i.e., \dred{} with minimal subtraction) is to be a viable
renormalization scheme, then one should be able to transform physical
results from one scheme into the other by finite shifts of the
renormalized parameters. This property has been confirmed several
times~\cite{Jack:1994bn,Harlander:2006rj,Harlander:2006xq}.  The proper
conversion relation for the strong
coupling between the $\msbar$ and the $\drbar$ scheme in $n_f$-flavor
standard \qcd{} is given at two-loop level by~\cite{Harlander:2006rj}
\begin{equation}
\begin{split}
  \asMS{} &= \asDR{}\left[1-\frac{\asDR{}}{4\pi}
    - \frac{5}{4}\left(\frac{\asDR{}}{\pi}\right)^2
    + \frac{\asDR{}\alpha_e}{12\pi^2}\,n_f+\ldots\right]\,,
  \label{eq::asMS2DR_2}
\end{split}
\end{equation}
where $\asMS$ denotes the strong coupling in the $\msbar{}$ scheme.
Three-loop corrections to this relation are known as
well~\cite{Harlander:2006xq}.

When evaluating physical observables in \dreg{}, the result depends only
on $\asMS$, while it depends on both $\asDR$ and $\alpha_e$ in \dred{}.
This ambiguity should be viewed as a freedom of the renormalization
scheme: any choice of $\aeDR$
determines the value of $\asDR{}$ by comparison to the experimental
value of the physical observable at one particular scale $\mu_0$. At any
other scale $\mu$, $\asDR$ and $\alpha_e$ are determined by the {\abbrev
RGE}s \eqn{eq::rges}.

 Also, there is a
unique relation between the perturbative coefficients of the $\drbar{}$
and the $\msbar{}$ expression of the physical observable, to be called
$R$ and $\bar R$ in what follows.  For example, assume that
\begin{equation}
\begin{split}
R(\asDR,\aeDR) &=
\sum_{i,j \geq 0} \left(\frac{\asDR}{\pi}\right)^i
\left(\frac{\alpha_e}{\pi}\right)^j\,r_{ij}\,,\qquad\qquad
\bar R(\asMS) = \sum_{i\geq 0} \left(\frac{\asMS}{\pi}\right)^i\, \bar r_i\,.
 \label{eq::RMS}
\end{split}
\end{equation}
Then, inserting Eq.\,(\ref{eq::asMS2DR_2}) into Eq.\,(\ref{eq::RMS}) and
requiring equality, one derives the relations
\begin{equation}
\begin{split}
r_{00} = \bar r_0\,,\qquad
r_{10} = \bar r_1\,,\qquad
r_{01} =  0\,,\qquad
r_{20} = \bar r_2 - \frac{\bar r_1}{4}\,,\qquad
r_{02} = 0\,,\qquad
r_{11} =  0\,,\\
r_{30} = \bar r_3 -\frac{\bar r_2}{2} - \frac{5}{4}\,\bar r_1\,,\qquad
r_{21} = \frac{n_f}{12}\,\bar r_1\,,\qquad
r_{12} = 0\,,\qquad
r_{03} = 0\,,\qquad \mbox{etc.}
\end{split}
\end{equation}

\section{Relation of \boldmath{$\alpha_s$} and \boldmath{$\alpha_e$} by Supersymmetry}

Supersymmetry is a concept that provides solutions to some of the most
pressing questions left open by the Standard Model. As already mentioned
above, in a \susy{} theory it is required that $\asDR{}=\alpha_e$ at all
energy scales, and thus $\beta_s=\beta_e$. We can use the \qcd{} results
of Ref.\,\cite{Harlander:2006rj} to test the consistency of \dred{} and
\susy{} for a \susy{} Yang Mills theory at three-loop level, simply by
choosing the color factors appropriately. Indeed, we find
that $\beta_s=\beta_e$ through three loops in a \susy{} Yang Mills
theory. For a check of this relation within \susy{}-\qcd{}, one needs to
include chiral fields in the fundamental representation of the gauge
group, or in other words, quarks and squarks. This is work in progress.

If indeed the \qcd{} that we observe is the low energy limit of a softly
broken \susy{}-\qcd{} theory, then the freedom of choosing $\alpha_e$ is
lost, because within this \susy{} theory, we require $\alpha_e^{\rm
(full)}=\asDR^{\rm (full)}$ at all scales.  The couplings in \qcd{} are
related to those in \susy{}-\qcd{} by matching relations:
\begin{equation}
\begin{split}
\asDR(\mu) = \zeta_s \asDR^{\rm (full)}(\mu)\,,\qquad
\alpha_e(\mu) = \zeta_e \alpha_e^{\rm (full)}(\mu)\,,
\label{eq::decoupling}
\end{split}
\end{equation}
where $\zeta_s$ and $\zeta_e$ are functions of $\asDR^{\rm (full)}$, the
\susy{} particle masses, and the ``matching scale'' $\mu$ (if $\asDR$
and $\alpha_e$ are the couplings in five-flavor \qcd{}, then $\zeta_s$
and $\zeta_e$ depend also on the top quark mass). Note that since the
dependence of $\zeta_{s,e}$ on the matching scale $\mu$ is logarithmic,
one should apply \eqn{eq::decoupling} at a scale not too much different
from the \susy{} particle masses. Also, if these masses are spread over
a large range, matching better be done in several steps.

$\zeta_s$ and $\zeta_e$ can be evaluated perturbatively. The two-loop
expression for $\zeta_s$ has been calculated in
Ref.\,\cite{Harlander:2005wm}, while for $\zeta_e$ only the one-loop
term is known~\cite{Harlander:phen}.

Assume now for the sake of the argument that all \susy{}-\qcd{} particle
masses are identical, say $m_{\tilde q} = m_{\tilde g} = \tilde M \sim
1$\,TeV. 
If $\asMS(M_Z)$ in \qcd{} is given by experiment, then the
\susy{} coupling, for example at the GUT scale $\mu_{\rm GUT}$, can be
determined by the following scheme:
\begin{equation}
\begin{split}
\asMS(M_Z)
\stackrel{(i)}{\rightarrow}
\asMS(\mudec)
\stackrel{(iii)}{\leftarrow}
\left\{
\begin{array}{c}
\asDR(\mudec) \\
\alpha_e(\mudec)
\end{array}
\right\}
\stackrel{(ii)}{\leftarrow}
\asDR^{\rm (full)}(\mudec)
\stackrel{(iv)}{\rightarrow}
\asDR^{\rm (full)}(\mu_{\rm GUT})\,.
\end{split}
\end{equation}
If the evolution is to be consistent through $n$-loop order, then steps
$(i)$ and $(iv)$ need to be done through $n$ loops, while steps $(ii)$
and $(iii)$ are only required through $(n-1)$ loops.  Here, it is
understood that one starts with a trial value $\alpha_0$ for $\asDR^{\rm
(full)}(\mudec)$, evaluates steps $(ii)$ and $(iii)$, and compares the
value for $\asMS(\mudec)$ obtained in this way with the one obtained
from step $(i)$. If it agrees, one performs step $(iv)$ with $\asDR^{\rm
(full)}(\mudec) = \alpha_0$, otherwise, one starts again with a
different value for $\alpha_0$.

\begin{wrapfigure}{r}{0.5\columnwidth}
\centerline{\includegraphics[width=0.55\columnwidth]{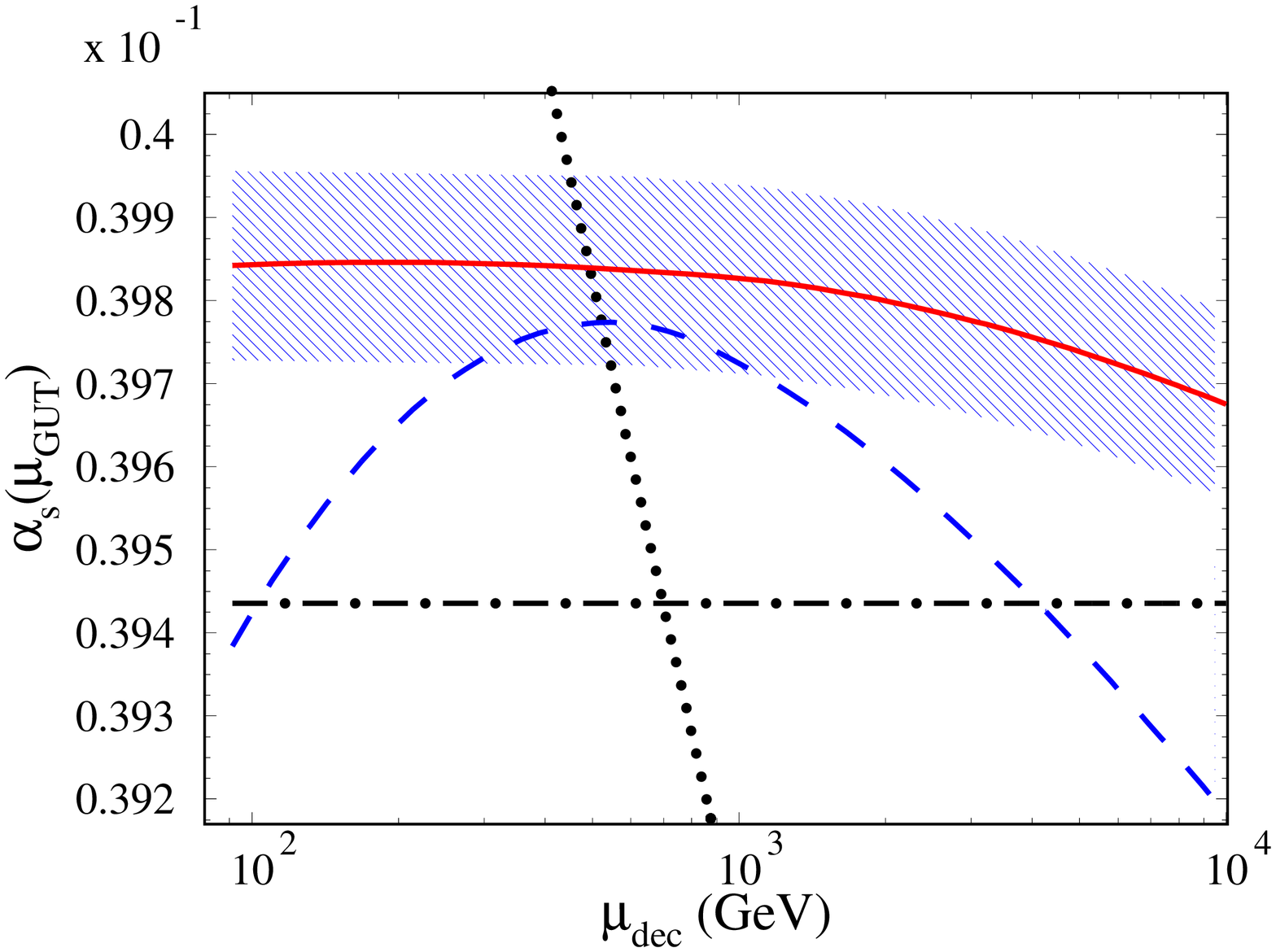}}
\caption{$\alpha_s$ at $\mu_{\rm GUT}\equiv 10^{16}$\,GeV derived from
  $\alpha_s(M_Z)$ in 1-, 2-, and 3-loop approximation (dotted, dashed,
  solid) as a function of the decoupling scale $\mudec$. The dash-dotted
  curve is what results from the formula given in
  Ref.\,\cite{Aguilar-Saavedra:2005pw}. See Ref.\,\cite{Harlander:phen}
  for details.}\label{fig::asgut}
\end{wrapfigure}
An alternative way to proceed was applied in
Ref.\,\cite{Harlander:phen}. There, the relation between $\asDR(\mudec)$
and $\alpha_e(\mudec)$ was perturbatively expanded such that
$\asDR(\mu_{\rm GUT})$ could be directly evaluated from $\asMS(M_Z)$
without the need for an iterative procedure. The difference between
these two approaches is formally of higher orders in $\alpha_s$, but is
expected to grow as the decoupling scale moves away from the \susy{}
masses $\tilde M$. At three-loop level, the two approaches are
consistent with each other within each others uncertainty (derived from
the experimental error on $\alpha_s(M_Z)$~\cite{Bethke:2006ac}) over a
large range of the decoupling scale. Figure\,\ref{fig::asgut} shows the
result~\cite{Harlander:phen}, demonstrating the numerical importance of
the three-loop effects, in particular if decoupling is done at other
scales than $\tilde M$ (quite often one finds $\mudec = M_Z$, for
example~\cite{Aguilar-Saavedra:2005pw}). 

\section{Conclusions}

\dred{} is currently considered the appropriate regularization method
for supersymmetric theories. Applied to non-\susy{} theories, it leads
to evanescent couplings, with their own evolution and decoupling
relations. If parameters from the non-\susy{} theory are to be related
to \susy{} parameters, the conversion relations will typically involve
these evanescent couplings. Here we took these issues into account for
the derivation of $\alpha_s(\mu_{\rm GUT})$ from $\alpha_s(M_Z)$ at
three-loop level.

\begin{footnotesize}

\end{footnotesize}

\end{document}